\documentclass[
 reprint,
 amsmath,amssymb,
 aps,
]{revtex4-2}

\usepackage{graphicx}
\usepackage{dcolumn}
\usepackage{bm}
\usepackage{hyperref}
\usepackage{mathrsfs}
\usepackage{amsmath}
\usepackage{tikz}
\usepackage{subcaption}
\usepackage[compatibility=false]{caption}
\usepackage{ragged2e}
\begin{document}

\title{Logarithmic growth of operator entanglement in a clean non-integrable circuit}

\author{Mao Tian Tan} \affiliation{Faculty of Mathematics and
  Physics, University of Ljubljana, Jadranska 19, SI-1000 Ljubljana,
  Slovenia}

\author{Toma\v z Prosen} \affiliation{Faculty of Mathematics and
  Physics, University of Ljubljana, Jadranska 19, SI-1000 Ljubljana,
  Slovenia} \affiliation{Institute of Mathematics, Physics and
  Mechanics, Jadranska 19, SI-1000 Ljubljana, Slovenia}

\date{\today}

\begin{abstract}
We study a so-called semi-ergodic brickwork dual-unitary circuits where, in the infinite volume limit, the two-point correlation functions of single-site operators exhibit ergodic behavior along one light ray and non-ergodic behavior along the other light ray. Here, however, we study intermediate and long-time dynamics of a system in a finite, large volume. Under such dynamics, the Heisenberg evolution of a single traceless single-site operator lies within a restricted subspace, and this time evolution can be mapped to a simpler problem of a single qutrit scattering with a bunch of qubits sequentially. Despite the model being non-integrable and free from any quenched disorder, the operator entanglement grows at most logarithmic in time, contrary to prior expectations. 
The auto-correlation function can be written in terms of a sum of products of $SO(3)$ matrices, allowing for a random matrix prediction for the auto-correlation function at late times. The operator size distribution also becomes bimodal at certain times, displaying intermediate behavior between chaotic and free systems.
\end{abstract}

\maketitle


\section{Introduction}
Despite the remarkable advances in quantum computing and quantum simulators \cite{Bernien2017,Chertkov2022,Fischer2026,Lau2022}, a clear and unambiguous demonstration that quantum computers can outperform classical computers in practical and useful real-world tasks has yet to be achieved. This challenge goes by the name of quantum advantage \cite{eisert2025mind,huang2025vast,lanes2025framework,haghshenas2025digital,Arute2019,PhysRevLett.127.180501,PhysRevX.15.021052,abanin2025constructive,PRXQuantum.5.010308,tindall2025dynamics}, and its resolution is central to the emerging field of quantum technology.

There are various classes of problems that are potential candidates for the demonstration of quantum advantage, among which is the simulation of the time evolution of quantum mechanical systems. Quantum dynamics is notoriously difficult to simulate on classical computers because of the rapid generation of entanglement entropy and is thus challenging to access using common classical computational techniques such as tensor networks \cite{ORUS2014117}. In particular, we would expect quantum chaotic systems to be difficult to simulate due to their intractable dynamics and rapid entanglement generation. The issue with chaotic dynamics is that correlation functions decay exponentially in time in such systems, making these observables difficult to measure on a quantum computer. Integrable systems, on the other hand, can have non-decaying or much slower decaying time correlation functions that can be measured on a quantum computer, but are expected to be easy to simulate, at least in principle, and in some cases can be solved analytically, due to the presence of a large number of symmetries.

In this work, we seek to explore a new kind of dynamics that sits somewhere between these two regimes of chaoticity and integrability. A convenient family of quantum mechanical systems that can realize such dynamics are the dual-unitary circuits \cite{PhysRevLett.123.210601,Bertini2021,PhysRevX.9.021033,PhysRevResearch.2.033032,PhysRevLett.126.100603,PhysRevResearch.7.L012011,bertini2025exactly,PhysRevLett.130.090601,PhysRevLett.125.070501,PhysRevResearch.3.043046,Fischer2026,Chertkov2022,Claeys2022emergentquantum,PhysRevB.106.L201104,Kasim_2023,PhysRevB.102.174307,PhysRevB.101.094304,10.21468/SciPostPhys.16.2.049,Suzuki2022computationalpower,Song2025montecarlo,PhysRevB.111.024301}. The original brickwork dual-unitary circuits have two-point correlation functions between traceless, single-site operators that are non-vanishing only on two light rays emanating from one of these operators. In brickwork dual-unitary circuits, two-point correlation functions along each light ray are controlled by independent quantum channels, and by a judicious choice of the dual-unitary circuit, we can have ergodic dynamics of two-point correlations along one light ray and non-ergodic dynamics along the other. This gives rise to a kind of intermediate dynamics we dub \emph{semi-ergodic} dynamics which could potentially be too chaotic to be classically simulatable and integrable enough for correlation functions to survive at late times, beyond the reach of classical numerical methods. To prevent analytic tractability, we consider brickwork circuits of finite widths~\footnote{In the infinite width case this kind of dynamics has been studied in a different context in Ref.~\cite{10.21468/SciPostPhys.8.4.068}} since these cannot be contracted using standard diagrammatic approaches beyond early times. Since these circuits are not integrable and their gates have a finite entangling power, we expect a priori the operator entanglement to grow linearly, a signature of classical simulation complexity~\cite{ProsenZnidaric2007,PhysRevA.76.032316}.

However, surprisingly, the semi-ergodic dual-unitary circuit we consider in this paper exhibits logarithmic operator entanglement growth, providing a novel example of a non-integrable system that has slow growth of operator entanglement. It also exhibits intriguing dynamical properties that are in between what one would expect for chaotic and integrable systems, such as a bimodal distribution in the operator size for a single site initial Pauli operator undergoing Heisenberg evolution. While this model is insufficient to demonstrate quantum advantage, we hope that these results spur further investigation into semi-ergodic dynamics as a possible approach to complexity classification of quantum dynamics, and to uncover its novel dynamical properties which can enrich our understanding on the interplay between chaotic and integrable systems.

We begin by describing the setup in section \ref{Setup}. We show that the Heisenberg evolution of a single-site traceless operator undergoing this semi-ergodic dual-unitary dynamics in a finite periodic brickwork circuit can be mapped into a single qutrit which scatters sequentially with a bunch of qubits which wind around a helix. These qubit states can be thought of as encoding the presence or absence of a soliton. By utilizing a hidden symmetry in this dynamics, we rewrite the correlation functions in terms of products of three-dimensional orthogonal matrices, and we argue that the late-time autocorrelation functions are well-described by random matrix theory. In section \ref{NumericalResults}, we present numerical results of the operator entanglement, auto-correlation functions and the operator size distribution. We conclude with a discussion and a vision for future research directions.

\section{Semi-Ergodic Dual Unitary Circuits}\label{Setup}
We begin this section by describing the quantum circuit that we are studying. Then, we explain how the Heisenberg dynamics of a local Pauli operator under the evolution by this quantum circuit can be described by a bunch of non-interacting qubits that interact only with a single qutrit, one at a time. The expressions for the correlation functions are simplified analytically, allowing us to make some asymptotic predictions based on random matrix theory. Various quantities such as the operator size and operator entanglement are computed numerically and the data are presented.
\subsection{Setup and definitions}
Construct a brickwork circuit acting on a chain of $L$ qubits labelled $1,2,\ldots,L$ as follows. Define the right translation operator as 
\begin{equation}
    \Pi |a_1 a_2 \ldots a_L\rangle = |a_L a_1 a_2 \ldots a_{L-1}\rangle
\end{equation}
where $a_i\in\{0,1\}$ for $i=1,2,\ldots,L$ which labels the computational (qubit) basis. Impose periodic boundary conditions so that site $L+1$ is identified with site 1. In this paper, we take the number of sites $L$ to be even and the Floquet operator can be written as
\begin{align}\label{FloquetUnitary}
    \mathscr{U}_F =& \mathscr{U}_e\mathscr{U}_o, \nonumber\\ \mathscr{U}_o=&U^{\otimes L/2} ,\nonumber \\ \mathscr{U}_e =&\Pi^{-1} \mathscr{U}_o \Pi
\end{align}
so the full brickwork circuit can be obtained by repeated applications of $\mathscr{U}_F$. Denote the circuit with $t$ layers of either $\mathscr{U}_e$ or $\mathscr{U}_o$ with $\mathscr{U}(t)$\footnote{In particular, $\mathscr{U}(2)=\mathscr{U}_F$}. Typical brickwork circuits are obtained by raising the Floquet unitary to some power and thus contain even layers of unitaries but here we relax this condition and allow for either odd or even total number of layers of unitaries, as long as the layers $\mathscr{U}_e$ and $\mathscr{U}_o$ are applied in an alternating fashion with the first layer being $\mathscr{U}_o$.

Next, we specialize the gate $U$ to be dual-unitary which for qubit systems can be generally written, in appropriate gauge, as \cite{PhysRevLett.123.210601}
\begin{equation}\label{DUMatrix}
    U = V\left[\frac{\pi}{4},\frac{\pi}{4},J\right] (v_-\otimes v_+)
\end{equation}
where we define the two-qubit Heisenberg gate
\begin{eqnarray}
    &&V[J_x,J_y,J_z]\equiv \\ &&\quad\exp\left[-i \left(J_x\sigma^x\otimes \sigma^x+J_y\sigma^y\otimes \sigma^y+J_z\sigma^z\otimes \sigma^z\right)\right] \nonumber
\end{eqnarray}
and $v_{\pm}$ are two single-qubit unitary gates.
As explained in \cite{PhysRevLett.123.210601}, the infinite temperature two-point correlation function of two single-site traceless operators vanish identically unless they are null-like separated. In the latter case, the rate of decay of the two-point correlation function is determined by two quantum channels $\mathcal{M}_+$ and $\mathcal{M}_-$ which for our parametrization of the two-site dual-unitary gate in \eqref{DUMatrix} are determined by $v_+$ and $v_-$ respectively, as well as $J$. The parameter $J$ determines the entangling power of the two-site dual-unitary \cite{PhysRevLett.125.070501,PhysRevA.62.030301}, with $J=0$ corresponding to maximal entangling power and $J=\frac{\pi}{4}$, at which point $V$ becomes a SWAP gate, has no entangling power \cite{PhysRevLett.123.210601,PhysRevA.69.032315}. The two-point correlation function for two null-like separated single-site traceless operators $a$ and $b$ for brickwork dual-unitary circuits simplify to, $\nu\in\{\pm\}$:
\begin{equation}
    \frac{1}{2^L}\text{Tr}\left[b_{x+2t\nu}\mathscr{U}(2t)^\dagger a_x \mathscr{U}(2t)\right] = \frac{1}{2}\text{Tr}\left[\mathcal{M}_\nu^{2t}(a)b\right],
\end{equation}
The quantum channel $\mathcal{M}_+$ describes the correlation functions where the two single-site operators are seperated by a right-moving lightray while the quantum channel $\mathcal{M}_-$ describes the correlation functions where they are separated by a left-moving lightray.
Because these are unital quantum channels, the identity is always an eigenoperator with eigenvalues 1. The channels are called non-ergodic if there is at least one non-trivial operator with eigenvalue 1, and ergodic otherwise. In the former case, the two-point correlation functions of single-site traceless operators can survive at late times (however, only in the infinite width limit $L\to\infty$) while in the latter case such correlation functions will decay to zero.

\subsection{Semi-ergodic unitary evolution}

Take $\mathcal{M}_+$ to be non-ergodic so that $v_+$ is just a combination of $I$ and $\sigma^z$. For the subsequent time evolution of a single-site Pauli matrix located on an odd site, its parameters completely drop out of the evolution as we shall see. Parametrize the ergodic single-site unitary as 
\begin{equation}\label{SingleSiteErgodic}
    v_- = \begin{pmatrix}
        e^{i\psi} \cos{\theta}& e^{i\phi} \sin{\theta} \\
        -e^{-i\phi} \sin{\theta}& e^{-i\psi} \cos{\theta}
    \end{pmatrix}
\end{equation}
For generic choices of Euler angles $\theta$, $\psi$ and $\phi$, the corresponding quantum channel $\mathcal{M}_-$ will be ergodic. Because the quantum channel $\mathcal{M}_+$ is non-ergodic, a single-site Pauli matrix initially located at site 1 undergoing Heisenberg time evolution will not be a superposition of all $4^L-1$ traceless Pauli strings but will instead lie in an operator subspace of spanned by Pauli strings of the form 
\begin{equation}\label{PauliStringsNonZeroCorrelatorSpecialForm}
    \sigma_{1-t}^{\alpha} \prod\limits_{\substack{j-(1-t)\equiv 0 \,\text{mod 2}\,\\ j\neq 1-t}} \mathbb{I}_j \prod_{k-(1-t)\equiv 1\,\text{mod}\,2} \sigma_k^{\beta_k}
\end{equation}
where $t$ is the number of layers (not Floquet operators $\mathscr{U}_F$ in \eqref{FloquetUnitary} but individual layers $\mathscr{U}_e$ and $\mathscr{U}_o$) of unitaries applied. Here, $\alpha=1,2,3$ and $\beta_k = 0,3$ for $k=1,2,\ldots,L/2$. This is a $3\times 2^{L/2}$ dimensional invariant space of operators which is smaller than the entire operator Hilbert space by an exponential factor. By the two-site translational invariance of the brickwork circuit $\mathscr{U}(t)$, we could have placed the initial Pauli matrix on any other odd site but we fix it to be site 1 for concreteness. It is straightforward to check that conjugating \eqref{PauliStringsNonZeroCorrelatorSpecialForm} with $\mathscr{U}_e$ in \eqref{FloquetUnitary} maps it to strings of the same form but with $t\rightarrow t+1$, i.e. the invariant operator space is then generated by Pauli strings of the form \eqref{PauliStringsNonZeroCorrelatorSpecialForm} but shifted one site to the left. Conjugating the resulting Pauli strings with $\mathscr{U}_o$ leads to an invariant operator space generated by strings of the form \eqref{PauliStringsNonZeroCorrelatorSpecialForm} but with $t\rightarrow t+2$. This is diagramatically represented as

\begin{align}
    &\begin{tikzpicture}[x=1.2cm,y=1cm]
  \tikzset{
    gate/.style={draw=blue!60!black, fill=blue!20, rounded corners=2pt, thick}
  }
  \def\h{1.2}
  \def\w{1.3}
  \def\s{0.4} 
  \def\xmin{0.0}
  \def\xmax{2.5*\w + 3*\s+0.5}
  \begin{scope}
    \clip (\xmin,-1) rectangle (\xmax,2);
    \draw[gate] (-0.5*\w,0) rectangle (0.5*\w,\h);
    \draw[gate] (0.5*\w+\s,0) rectangle (0.5*\w+\s+\w,\h);
    \node at (0.5*\w+\s+0.5*\w,\h/2) {$W$};
    \draw[gate] (0.5*\w+2*\s+\w,0) rectangle (0.5*\w+2*\s+2*\w,\h);
    \node at (0.5*\w+2*\s+1.5*\w,\h/2) {$W$};
    \draw[gate] (0.5*\w+3*\s+2*\w,0) rectangle (0.5*\w+3*\s+3*\w,\h);
\tikzset{leg/.style={draw=blue!60!black, fill=white, thick}}
\def\r{0.35}        
\def\y{-.35}         
\def\xL{0.5*\w-0.2}                         
\def\xA{0.5*\w+\s+0.2}                       
\def\xB{0.5*\w+\s+\w-0.2}                    
\def\xC{0.5*\w+2*\s+\w+0.2}                  
\def\xD{0.5*\w+2*\s+2*\w-0.2}                
\def\xR{0.5*\w+3*\s+2*\w+0.2}              
\draw[leg] (\xL,\y) circle (\r);
\node at (\xL,\y) {\scriptsize $I$};
\draw[leg] (\xA,\y) circle (\r);
\node at (\xA,\y) {\scriptsize $\sigma^{\beta_\frac{L}{2}}$};
\draw[leg] (\xB,\y) circle (\r);
\node at (\xB,\y) {\scriptsize $\sigma^\alpha$};
\draw[leg] (\xC,\y) circle (\r);
\node at (\xC,\y) {\scriptsize $\sigma^{\beta_1}$};
\draw[leg] (\xD,\y) circle (\r);
\node at (\xD,\y) {\small $I$};
\draw[leg] (\xR,\y) circle (\r);
\node at (\xR,\y) {\small $\sigma^{\beta_2}$};
  \end{scope}
\end{tikzpicture} \nonumber\\
&\hspace{2.5cm}\Downarrow\\
&\begin{tikzpicture}[x=1.2cm,y=1cm]
  \tikzset{
    leg/.style={draw=blue!60!black, fill=white, thick},
    wire/.style={thick}
  }
  \def\h{1.2}
  \def\w{1.3}
  \def\s{0.4}
  \def\r{0.35}
  \def\yTop{0.35}
  \def\xL{0.5*\w-0.2}
  \def\xA{0.5*\w+\s+0.2}
  \def\xB{0.5*\w+\s+\w-0.2}
  \def\xC{0.5*\w+2*\s+\w+0.2}
  \def\xD{0.5*\w+2*\s+2*\w-0.2}
  \def\xR{0.5*\w+3*\s+2*\w+0.2}
  \foreach \x/\lab in {
    \xL/{\scriptsize $\sigma^{\beta_{\scalebox{0.7}{$\frac{L}{2}-1$}}}$},
    \xA/{\scriptsize $\sigma^{a}$},
    \xB/{\scriptsize $\sigma^b$},
    \xC/{\scriptsize $I$},
    \xD/{\small $\sigma^{\beta_1}$},
    \xR/{\small $I$}
  }{
    \draw[wire] (\x,0) -- (\x,\h);           
    \draw[leg]  (\x,\h+\yTop) circle (\r);   
    \node at    (\x,\h+\yTop) {\lab};        
  }
\end{tikzpicture}
\nonumber
\end{align}


\noindent where $W = U\otimes U^*$ is the folded two-site gate and the vertical lines stand for the folded identity. Here, $\beta_1,\ldots\beta_L,b\in\{0,3\}$ and $\alpha,a\in\{1,2,3\}$. We can think of $\alpha,a$ as labeling a qutrit (whose three states correspond to $\sigma^x$, $\sigma^y$ and $\sigma^z$) that is traveling in the ergodic direction, corresponding to the ergodic channel $\mathcal{M}_-$,
while $\beta_1,\ldots\beta_{L/2},b$ label $L/2$ qubits (whose two states correspond to $I$ or $\sigma^z$) traveling in the non-ergodic direction corresponding to $\mathcal{M}_+$. Note that the only non-trivial evolution occurs where the unitaries act on $\sigma^{\beta_L}\otimes \sigma^\alpha$. All the other qubits simply move along the light ray undisturbed.
\begin{figure}
    \centering
    \begin{tikzpicture}[line cap=round,line join=round]

\def\W{4.8}     
\def\H{13.7}    
\def\s{1.6}     

\definecolor{myblue}{RGB}{0,102,255}
\definecolor{myred}{RGB}{220,50,32}
\definecolor{myyellow}{RGB}{220,220,0}
\definecolor{myorange}{RGB}{255,170,0}

\draw[myorange] (0,0) -- (0,0.7*\H);
\draw[myorange] (\W,0) -- (\W,0.7*\H);

\begin{scope}
\clip (0,0) rectangle (\W,0.7*\H);

\foreach \k in {-3,...,12}{
\draw[myblue,line width=1.1pt]
(-0.5*\s,{\k*\s}) -- (\W,{(\k*\s)+\W+0.5*\s});
}

\foreach \k in {-2,...,12}{
\draw[myyellow,line width=1.1pt]
(0,{(\k*\s)+\s}) -- (\W,{(\k*\s)+\s-\W});
}

\foreach \k in {2,5,8,11}{
\draw[myred,line width=1.1pt]
(0,{(\k*\s)+\s}) -- (\W,{(\k*\s)+\s-\W});
}

\foreach \k in {2,5,8}{
\foreach \x in {0.4,1.2,2,2.8,3.6,4.4}{
\fill (\x,{(\k*\s)-\x+\s}) circle (2.3pt);
}
}

\end{scope}

\draw[gray!70] (0,0.8) -- ({\W},0.8);
\draw[gray!70] (0,1.6) -- ({\W},1.6);
\draw[gray!70] (0,2.4) -- ({\W},2.4);

\node[left] at (-0.3,0.8) {\bfseries t=1};
\node[left] at (-0.3,1.6) {\bfseries t=2};

\draw[->] (-0.65,2.2) -- (-0.65,3.4);


\node[anchor=west,font=\bfseries] at (-2.5,6.1) {X, Y, Z};
\draw[myred,line width=1.2pt] (-2.8,6.3) -- (-1.4,4.9);

\node[anchor=west,font=\bfseries] at (-2.95,2.8) {I,Z};
\draw[myblue,line width=1.2pt] (-2.8,2.0) -- (-1.4,3.4);

\fill (-2.6,1.15) circle (2.8pt);
\node[anchor=west,font=\bfseries] at (-2.5,1.15) {=A};
\end{tikzpicture}
    \caption{The brickwork dual-unitary circuit with periodic boundaries generated by the Floquet unitary \eqref{FloquetUnitary} is mapped to a qutrit (red line) propagating in the ergodic direction and $\frac{L}{2}$ qubits (blue lines) propagating in the non-ergodic direction. The yellow lines are the identities in \eqref{PauliStringsNonZeroCorrelatorSpecialForm} and do not participate in the dynamics. The black dots correspond to the scattering between the qutrit and a single qubit at each time step, described by the 
    $6\times 6$ orthogonal matrix $A$ \eqref{SixBySixMatrix}, which corresponds to a single layer of unitaries $\mathscr{U}_e$ or $\mathscr{U}_o$.     The above diagram is drawn for $L=6$.}
    \label{QutritQubitScatteringDiagram}
\end{figure}
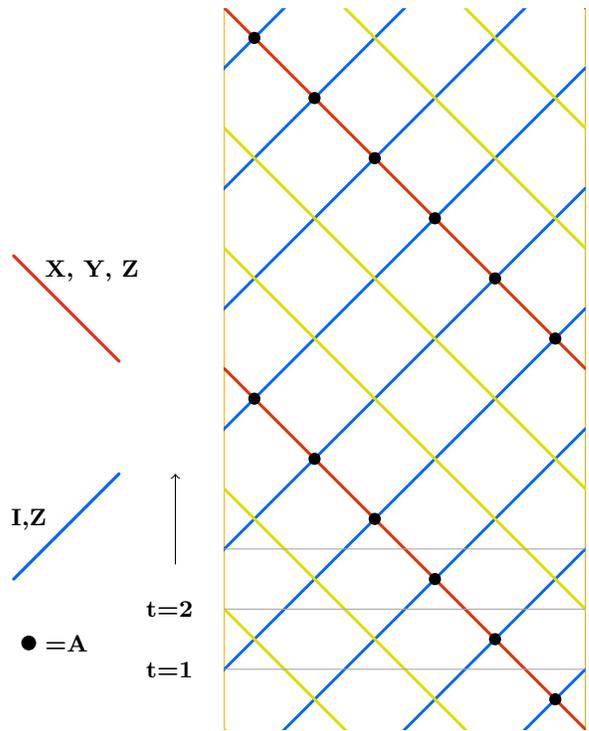

Since the initial $X$ or $Y$ operator evolves within an invariant subspace \eqref{PauliStringsNonZeroCorrelatorSpecialForm}, we can utilize this knowledge to more efficiently compute their correlation functions. Consider an operator lying in the span of Pauli strings in \eqref{PauliStringsNonZeroCorrelatorSpecialForm}
\begin{align}\label{PauliStringZBasis}
    O =\sum\limits_{\substack{\alpha=1,2,3 \\ \beta_1,\beta_2,\ldots,\beta_\frac{L}{2}=0,3}}& c_{\beta_1\beta_2\ldots\beta_\frac{L}{2}\alpha}\left(I\sigma^{\beta_j}\right)\ldots\left(I\sigma^{\beta_\frac{L}{2}}\right)  \nonumber\\\times&\left(\sigma^\alpha\sigma^{\beta_1}\right)\left(I\sigma^{\beta_2}\right)\ldots\left(I\sigma^{\beta_{j-1}}\right)
\end{align}
where parenthesis were drawn around sites $2i-1,2i$ for $i=1,\ldots,\frac{L}{2}$ for convenience. The coefficient $c_{\beta_1\beta_2\ldots\beta_\frac{L}{2}\alpha}$ is the operator wavefunction, and the very last index labels the left-moving qutrit $\sigma^\alpha$ where $\alpha=1,2,3$. The first index $\beta_1$ labels the diagonal operator $I$ or $\sigma^z$ immediately to the right of this $\sigma^\alpha$ while the second last index $\beta_\frac{L}{2}$ labels the diagonal operator $I$ or $Z$ immediately to the left of $\sigma^\alpha$. We do not label the spatial positions because the qutrit simply lies on the left-moving light ray propagating from the initial operator located at site 1, and the location of the other operators are determined relative to this qutrit.
Our semi-ergodic dual-unitary two-site gate \eqref{DUMatrix} satisfies the following properties
\begin{align}
    U^\dagger (\sigma^\beta\otimes I) U =& I \otimes \sigma^\beta,\nonumber \\ U^\dagger (\sigma^\beta\otimes\sigma^\alpha) U =& \sum\limits_{\substack{a=1,2,3\\ b=0,3}} A_{ab,\beta\alpha} \sigma^a \otimes \sigma^b
\end{align}
for $\alpha=1,2,3$ and $\beta=0,3$ and \begin{equation}\label{SixBySixMatrix}
    A_{ab,\beta\alpha} = \frac{1}{d^2} \text{Tr}\left((\sigma^a \otimes \sigma^b )U^\dagger (\sigma^\beta\otimes \sigma^\alpha )U\right)
\end{equation}
For the dual-unitary matrix \eqref{DUMatrix}, this $6\times 6$ real orthogonal matrix is a function of $J$, which parametrizes the two-site unitary $V$, as well as $v_-$, which determines the ergodic channel $\mathcal{M}_-$. This $6\times 6$ matrix describes the scattering of a qubit, $\sigma^\beta$ where $\beta=0,3$, with a qutrit, $\sigma^\alpha$ for $\alpha=1,2,3$. Note that the scattering process exchanges the position of the qutrit and the scattered qubit.

Evolving this operator with a single layer of such unitaries give
\begin{align}\label{SingleLayerEvolutionInvariantSubspace}
    &(U_{L1}U_{23},\ldots U_{L-2,L-1})^\dagger O (U_{L1}U_{23}\ldots U_{L-2,L-1})\nonumber\\
    =& \sum\limits_{\substack{a=1,2,3\\ b,\beta_1,\beta_2\,\ldots,\beta_{\frac{L}{2}-1}=0,3}}\bigg[ c'_{b\beta_1\ldots\beta_{\frac{L}{2}-1}a} (\sigma^{\beta_{j-1}}I)\ldots(\sigma^{\beta_{\frac{L}{2}-1}}\sigma^a)\nonumber\\
    \times&(\sigma^b I) (\sigma^{\beta_1}I)\ldots(\sigma^{\beta_{j-2}}I)\bigg]
\end{align}
where the new coefficients after conjugation by a single layer of unitaries $\mathscr{U}_e$ (not one Floquet time step of two layers but just a single layer) is 
\begin{equation}\label{OperatorWavefunctionUpdateRule}
    c'_{b\beta_1\ldots\beta_{\frac{L}{2}-1}a} = \sum\limits_{\substack{\alpha=1,2,3\\ \beta_{\frac{L}{2}}=0,3}} A_{ab,\beta_{\frac{L}{2}}\alpha}c_{\beta_1\beta_2\ldots\beta_\frac{L}{2}\alpha}
\end{equation}
This tells us how the coefficients for a single site operator initially located on an odd site are updated after conjugation by a single layer of unitaries $\mathscr{U}_e=U_{L1}U_{23}\ldots U_{L-2,L-1}$. Evolving the resulting strings with the next single layer of unitaries $\mathscr{U}_o=U_{12}U_{34}\ldots U_{L-1,L}$, as in the second line of \eqref{SingleLayerEvolutionInvariantSubspace}, gives the exact same update rule. This process is illustrated in figure \ref{QutritQubitScatteringDiagram}. Every time step corresponds to the application of a single layer of unitaries, either $\mathscr{U}_e$ or $\mathscr{U}_o$ in \eqref{FloquetUnitary}, which is equivalent to the scattering of the single qutrit with the qubit that is immediately to its left. All other qubits propagate freely until they scatter with that single qutrit. When the qubits are in the state that corresponds to the $\sigma^z$, they are essentially the solitons described in \cite{10.21468/SciPostPhys.8.4.068}. Therefore, figure \ref{QutritQubitScatteringDiagram} illustrates a qutrit interacting with a bath of solitions where each scattering process can create or annihilate a soliton.

\subsection{$U(1)$ symmetry of semi-ergodic DU gate}
The operator evolution in \eqref{OperatorWavefunctionUpdateRule} is written in the basis where $\sigma^{\beta_j}=I,Z$ in \eqref{PauliStringZBasis}. To consider the evolution of an operator in this invariant subspace in another basis, it is helpful to perform an operator state map
\begin{align}
    \sigma^\alpha \rightarrow& |\alpha\rangle\qquad\text{for}\qquad\alpha=1,2,3\nonumber\\
    \sigma^{\beta_j} \rightarrow&|\beta_j\rangle\qquad\text{for}\qquad \beta_j=0,3
\end{align}
for $j=1,2,\ldots,\frac{L}{2}$. The corresponding operator state for the  operator $O$ in \eqref{PauliStringZBasis} is a superposition of one qutrit and $\frac{L}{2}$ qubits.

Let us change the local qubit basis $\{|0\rangle,|3\rangle\}\rightarrow\{|0'\rangle,|1'\rangle\}$. Perform the change of basis 
\begin{equation}\label{ChangeOperatorStateBasis}
    |\nu\rangle = \sum_{\beta=0,3} V_{\beta \nu} |\beta\rangle
\end{equation}
for $\nu=0',1'$. The orthonormality of the new basis forces the matrix $V_{\beta \nu}$ to be unitary. 
Now, consider an operator in the invariant subspace \eqref{PauliStringZBasis}, but with the qubits expressed in this new basis,
\begin{align}\label{OperatorStateMapNewBasis}
    |O\rangle =\sum\limits_{\substack{\alpha=1,2,3\\ \nu_1,\nu_2,\ldots,\nu_{L/2}=0',1'}} f_{\nu_1,\nu_2,\ldots,\nu_{L/2}\alpha} |\nu_1 \nu_2\ldots \nu_{L/2}\alpha\rangle
\end{align}
where the coefficients $f_{\nu_1,\nu_2,\ldots,\nu_{L/2}\alpha}$ are linear combinations of the coefficients $c_{\beta_1\beta_2\ldots\beta_\frac{L}{2}\alpha}$ in \eqref{PauliStringZBasis}. In this basis, the operator state after a single layer of unitaries has been applied is
\begin{align}\label{OperatorStateXBasisSingleStepEvolution}
    |O'\rangle =& \sum\limits_{\substack{a,\alpha=1,2,3\\ b,\beta_{L/2}=0,3}} \sum_{\nu_1,\nu_2,\ldots,\nu_{L/2}} \bigg[A_{ab,\beta_{L/2}\alpha} f_{\nu_1,\nu_2,\ldots,\nu_{L/2}\alpha}  \nonumber\\
    \times&V_{\beta_{L/2}\nu_{L/2}} |b\nu_1\ldots\nu_{L/2-1} a\rangle\bigg]
\end{align}
It turns out that the six-by-six matrix $A$ has a $U(1)$ symmetry
\begin{equation}\label{U1Symmetry}
    A(e^{i\lambda \sigma^x}\otimes I_3) = (I_3\otimes e^{i\lambda \sigma^x}) A
\end{equation}
where $I_3$ is the identity acting on the qutrit Hilbert space. Since this $U(1)$ rotation is generated by $\sigma^x$, it is convenient to change our original basis in \eqref{PauliStringZBasis} to the eigenbasis of $\sigma^x$. Therefore, we set $|0'\rangle=(|0\rangle + |3\rangle)/\sqrt{2}\equiv|+\rangle$ and $|1'\rangle=(|0\rangle - |3\rangle)/\sqrt{2}\equiv|-\rangle$ so that the change of basis unitary in \eqref{ChangeOperatorStateBasis} becomes
\begin{equation}
    V = H \equiv (\vec{H}_+,\vec{H}_-)
\end{equation}
where $H$ is the Hadamard matrix. It is straightforward to show that for an arbitrary vector $\vec{u}\in\mathbb{C}^3$, \eqref{U1Symmetry} leads to
\begin{equation}
    A(\vec{H}_\nu \otimes \vec{u}) = (A_\nu \vec{u}) \otimes \vec{H}_\nu
\end{equation}
where $\nu=\pm$ and $A_{\pm}\in$ SO(3) and $[A_+,A_-]\neq0$ in general. We can apply this to \eqref{OperatorStateXBasisSingleStepEvolution} with $u_\alpha = f_{\nu_1\nu_2\ldots\nu_{L/2}\alpha}$ because for each summand, the indices $\nu_1\nu_2\ldots\nu_{L/2}$ are fixed. Hence, we find that the operator state after a single layer of unitaries is applied gets updated to 
\begin{align}\label{OperatorWavefunctionUpdateRuleXBasis}
    |O'\rangle =& \sum\limits_{\substack{a=1,2,3\\ \nu_1,\ldots,\nu_{L/2}=\pm}} \sum_{\alpha=1,2,3} \bigg[(A_{\nu_{L/2}})_{a\alpha} f_{\nu_1,\nu_2,\ldots,\nu_{L/2}\alpha} \nonumber \\
    \times&|\nu_{L/2} \nu_1\ldots \nu_{L/2-1}a\rangle\bigg]
\end{align}
So, if we expand the operator state in the $X$ basis, for each term $\nu_1,\ldots,\nu_{L/2}$, we simply multiply the three dimensional vector $\vec{f}_{\nu_1,\nu_2,\ldots,\nu_{L/2}}$ by an element of SO(3). Since the operator wavefunction evolution \eqref{OperatorWavefunctionUpdateRule} holds when the qutrit, whose position lies on the left-moving lightray emanating from the initial non-trivial single site operator, is on the odd or even lattice, \eqref{OperatorWavefunctionUpdateRuleXBasis} is true when the qutrit is on both the odd or even site as well. After applying a multiple of $\frac{L}{2}$ layers of unitaries, for $m\in \mathbb{N}$, we have
\begin{align}\label{AutoCorrelationFunction}
    &\langle O |O\left(m L/2\right) \rangle\nonumber\\&=\sum_{\nu_1\ldots \nu_{L/2}=\pm} \vec{f}_{\nu_1\ldots\nu_{L/2}}^\dagger\cdot \left(A_{\nu_1}\ldots A_{\nu_{L/2}}\right)^m \vec{f}_{\nu_1\ldots\nu_{L/2}}
\end{align}
This is the auto-correlation function of some operator $O$ after $m\frac{L}{2}$ single layers of unitary gates. At time steps which are multiples of $L/2$, the qutrit has scattered with every qubit the exact same number of times, and the evolution of the operator state has a periodic nature for each term in the expansion \eqref{AutoCorrelationFunction}.

The initial operator state is encoded in the three dimensional vectors $\vec{f}_{\nu_1\ldots\nu_{L/2}}$. Finally, we specialize to the case where $O$ is initially a single-site operator $O=\sum_\alpha a_\alpha \sigma^\alpha$. The operator state is 
\begin{equation}\label{SingleSitePauliOperatorState}
    |O\rangle = \sum_{\alpha=1,2,3}a_\alpha|0\ldots0,\alpha\rangle
\end{equation}
which is of the form \eqref{OperatorStateMapNewBasis} with $f_{\nu_1,\nu_2,\ldots,\nu_{L/2},\alpha}=\frac{1}{2^{L/4}}a_{\alpha}$. 

\section{Physical Observables}\label{NumericalResults}
In the previous section, we showed how a single-site Pauli matrix evolves under the semi-ergodic dual-unitary circuit generated by the Floquet unitary \eqref{FloquetUnitary}. In this section, we proceed to numerically study correlation functions, operator size and operator entanglement entropy.
Because the time evolution is given by multiplying a six-by-six matrix on a single Pauli vector, exact numerical calculations can be performed for relatively large system sizes (up to $L\approx 60$).

Empirically, the parameters $J$ and $\theta$ have the most significant effect on the correlation functions, so we will only explore the dependence of the following physical quantities on these two parameters. Unless otherwise specified, we set
\begin{equation}
    \psi = \frac{1+\sqrt{5}}{2}\frac{\pi}{2},\hspace{1cm} \phi = (1+\sqrt{2})\frac{\pi}{2}
\end{equation}

\subsection{Correlation Functions}
The simpest physical quantities to consider are the correlation function 
\begin{equation}
    C_{ab}(t) = \frac{1}{2^L} \text{Tr}(a\mathscr{U}^\dagger(t) b \mathscr{U}(t))
\end{equation}
where $\mathscr{U}(t)$ is the brickwork circuit containing $t$ layers of unitaries $\mathscr{U}_o$ or $\mathscr{U}_e$ in \eqref{FloquetUnitary}.

The auto-correlation function is given by a simple matrix product form \eqref{AutoCorrelationFunction}. When the number of such matrices applied is large, we expect, for generic $A_{\pm}\in SO(3)$, the product of matrices to converge to a Haar average so the auto-correlation function for a single-site Pauli matrix \eqref{SingleSitePauliOperatorState} should converge to, as $m\to\infty$
\begin{equation}\label{HaarAveragePrediction}
    \langle O |O\left(m L/2\right) \rangle \rightarrow \vec{a}^\dagger \int dU U \vec{a} = \vec{a}^\dagger \begin{pmatrix}
        1/3&&\\
        &1/3&\\
        &&1/3
    \end{pmatrix}\vec{a}
\end{equation}
where the integral is performed over the ensemble of Haar unitaries. Therefore, if the sum of $SO(3)$ matrices in \eqref{AutoCorrelationFunction} converges to an average over the Haar ensemble, we expect the auto-correlation to be given by the value $1/3$. To check if these auto-correlation functions have converged to the Haar-averaged result, we plot the finite-time averaged correlators
\begin{equation}\label{TimeAveragedCorrelator}
    \overline{C_{ab}}(t) \equiv \frac{1}{\frac{2t}{L}+1}\sum\limits_{\tau=0,\frac{L}{2},\ldots}^t C_{ab}(\tau)
\end{equation}
where we consider all the values of the correlators $\tau=0,\frac{L}{2},\ldots$ up to time $t$, which is a multiple of $\frac{L}{2}$, and we perform the average. 

\begin{figure}
\centering
\begin{subfigure}{0.42\columnwidth}
    \centering
    \includegraphics[height=3.8cm, keepaspectratio]{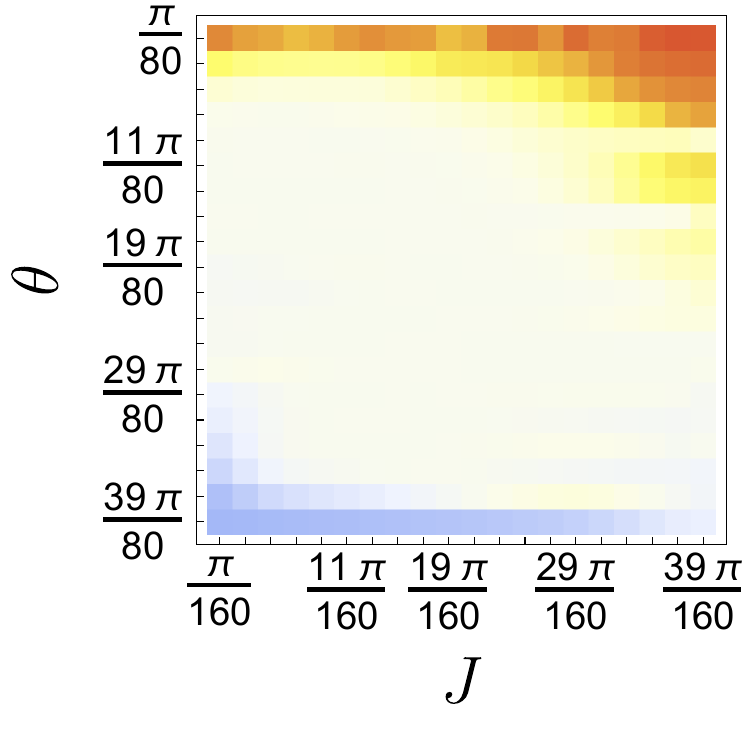}
    \caption{}
    \label{Autocorrelator_HeatMap:sub_a}
\end{subfigure}
\hfill
\begin{subfigure}{0.42\columnwidth}
    \centering
    \includegraphics[height=3.8cm, keepaspectratio,trim=1cm 0cm 0cm 0cm, clip]{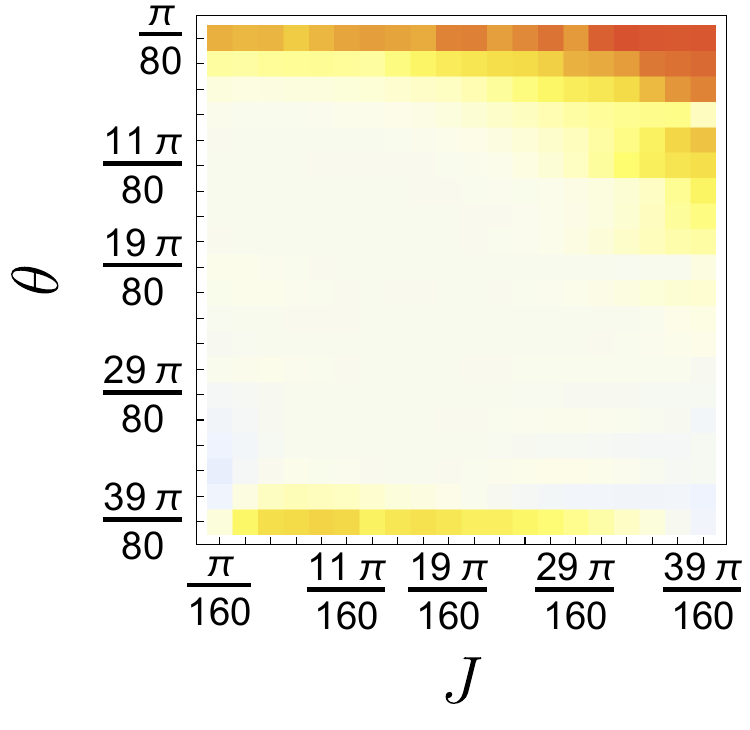}
    \caption{}
    \label{Autocorrelator_HeatMap:sub_b}
\end{subfigure}
\raisebox{1.5cm}{
\includegraphics[scale=0.5]{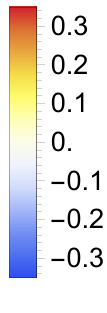}}
\caption{A heatmap of the difference between the Haar-averaged value and the time-averaged auto-correlator $\frac{1}{3}-\overline{C_{XX}}((\frac{L}{2})^2)$ \eqref{TimeAveragedCorrelator} evaluated at a time much larger than the system size $L$. The vertical axis is $\theta$ which parametrizes the single-site unitary $v_-$ \eqref{SingleSiteErgodic} while the horizontal axis is $J$ which parametrizes the entangling power of the two-site unitary. The plot on the left, (a), corresponds to $L=46$, while the plot on the right, (b), corresponds to $L=48$.}
\label{Autocorrelator_HeatMap}
\end{figure}

A heat map of $\overline{C_{XX}}((L/2)^2)$ is shown in figure \ref{Autocorrelator_HeatMap}. The auto-correlation functions of $\sigma^x$ appear to depend on the parity of $L/2$, so heat maps with $L/2$ odd are similar to each other and heat maps with $L/2$ even are similar to each other. For $L/2$ odd, the lowest part of the heat map is more negative than the corresponding points in the heat map where $L/2$ is even. For both odd and even $L/2$, the uppermost part of the heat map has large positive values, indicating that the product of $SO(3)$ matrices in \eqref{AutoCorrelationFunction} has not yet converged to the Haar averaged value, while large swathes of the heat map away from the uppermost and lowermost boundaries have small average values at $\mathcal{O}(L/2)$ multiples of $L/2$ time steps, indicating convergence of \eqref{AutoCorrelationFunction} to the Haar-averaged value. We refer to the parameters for which the autocorrelator has converged to the Haar-averaged prediction as semi-ergodic and the other points in parameter space as non-semi-ergodic. We note that this separation may well be system size dependent, and that in thermodynamic limit almost the entire parameter space 
may correspond to semi-ergodic regime. However, working at finite sizes, it appears that the non-semi-ergodic regions are quite stable against changing the system size $L$.

In the remainder of the paper, we shall compare two models for two specific parameter values, $\theta=\frac{21\pi}{80}$ and
$\theta=\frac{39\pi}{80}$, the former being semi-ergodic and the latter being non-semi-ergodic.
\begin{figure}
\centering
\begin{subfigure}{\columnwidth}
    \centering
    \includegraphics[width=\columnwidth]{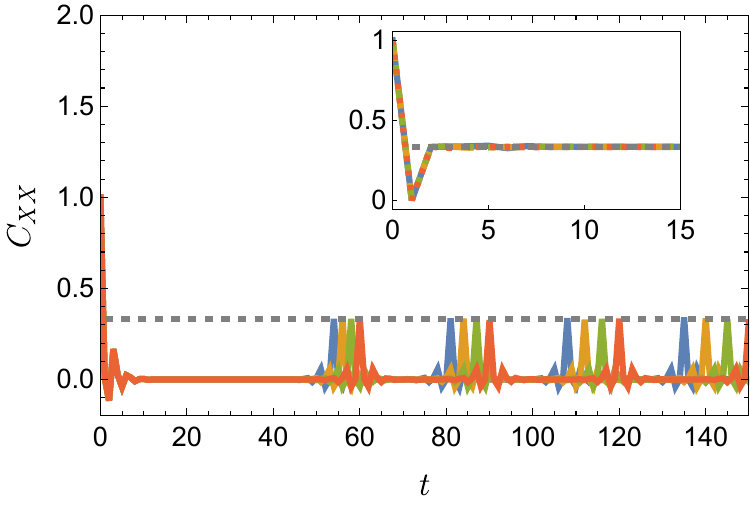}
    \caption{}
    \label{Autocorrelator:sub_a}
\end{subfigure}
\hfill
\begin{subfigure}{\columnwidth}
    \centering
    \includegraphics[width=\columnwidth]{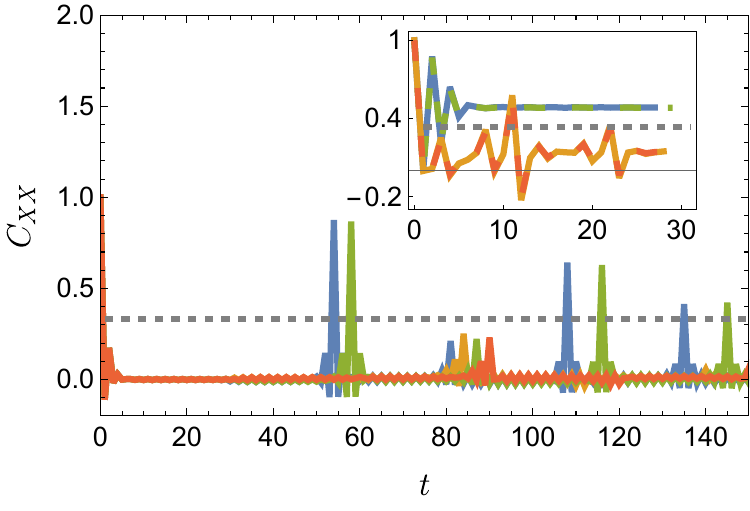}
    \caption{}
    \label{Autocorrelator:sub_b}
\end{subfigure}
\includegraphics[width=0.8\linewidth]{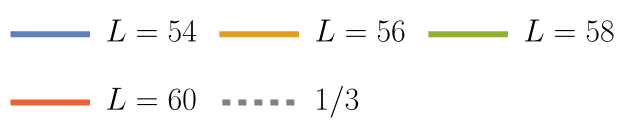}
\caption{(Main Plots) Plots of the auto-correlation functions $C_{XX}$ against the number of single layers of unitaries for $J=\frac{11\pi}{160}$, $\psi = \frac{1+\sqrt{5}}{2}\frac{\pi}{2}$, $\phi = (1+\sqrt{2})\frac{\pi}{2}$ and various system sizes $L$. The top plot (a) corresponds to a semi-ergodic point $\theta=\frac{21\pi}{80}$ while the bottom plot (b) corresponds to a non-semi-ergodic point $\theta=\frac{39\pi}{80}$. (Insets) The insets show the autocorrelators but only at time steps that are multiples of $L/2$. The horizontal axis is the number of multiples of $L/2$ unitaries that have been applied.}
\label{Autocorrelator}
\end{figure}

The autocorrelators are plotted in figure \ref{Autocorrelator} as a function of time. We see from the inset that in the semi-ergodic point of $\theta=\frac{21\pi}{80}$, the autocorrelator at time steps that are multiples of $L/2$ has converged to the Haar-averaged value of $1/3$. In fact, the auto-correlators at time steps of $L/2$ appears to be independent of $L$, indicating convergence in the system size. On the other hand, for the non-semi-ergodic point of $\theta=\frac{39\pi}{80}$, the autocorrelator at time steps of $L/2$, has not yet converged to the Haar-averaged value as expected from the heat map figure \ref{Autocorrelator_HeatMap}. The auto-correlator at time steps of $L/2$ almost appears to have converged in the system size as the plots for $L/2$ odd almost lie on top of each other and the plots for $L/2$ even almost lie on top of each other, as seen in the inset.

\subsection{Operator Entanglement}
\begin{figure}
\centering
\begin{subfigure}{\columnwidth}
    \centering
\includegraphics[width=\linewidth, keepaspectratio]{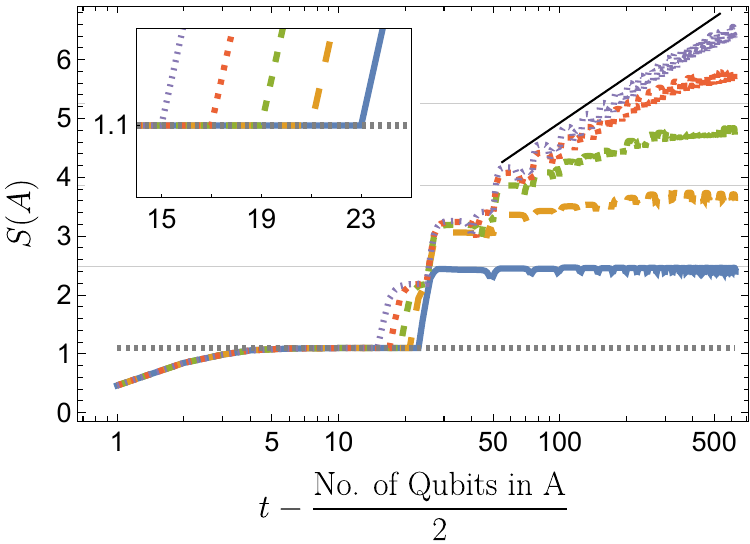}
    \caption{}
    \label{:sub_a}
\end{subfigure}
\begin{subfigure}{\columnwidth}
    \centering
\includegraphics[width=\linewidth, keepaspectratio]{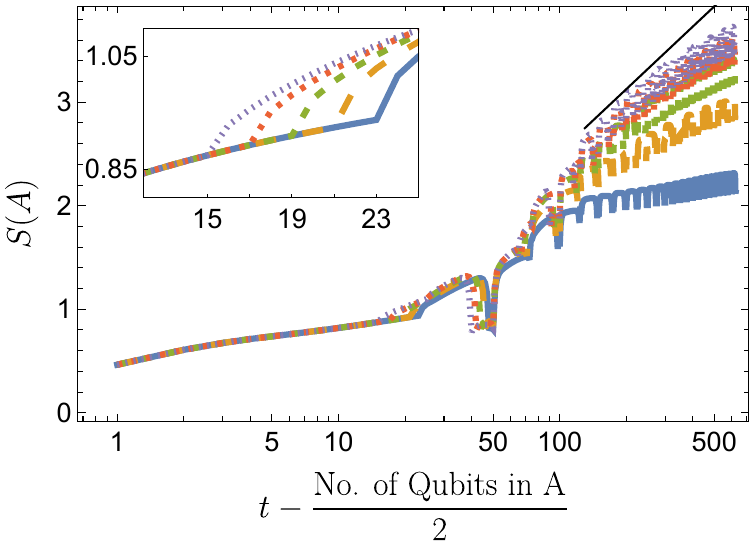}
    \caption{}
    \label{:sub_b}
\end{subfigure}
\includegraphics[scale=0.30]{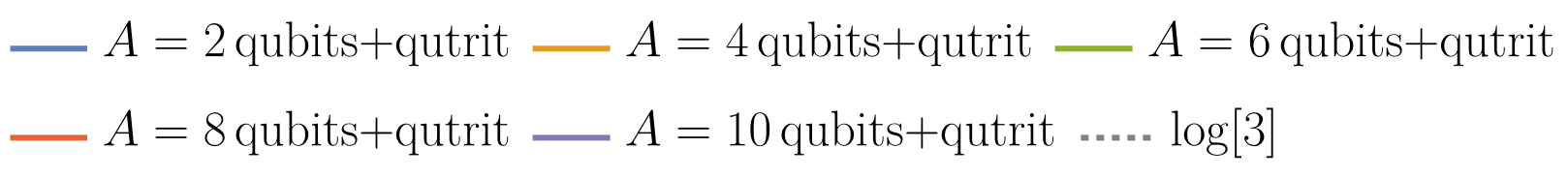}
\caption{Plots of the operator entanglement for a subsystem containing $l_A$ qubits and the one qutrit with an equal number of qubits on each side of the qutrit. The parameters are $J=\frac{11\pi}{160}$, $\psi = \frac{1+\sqrt{5}}{2}\frac{\pi}{2}$, $\phi = (1+\sqrt{2})\frac{\pi}{2}$ and the total system size is fixed at $L=50$. Plot (a) corresponds to the semi-ergodic point $\theta=\frac{21\pi}{80}$ while the plot (b) correspond to the non-semi-ergodic point $\theta=\frac{39\pi}{80}$. The plots are made with a logarithmic horizontal axes. The insets of these top two plots simply zoom in to the time when the different graphs start to diverge from one another. The x-axes for each graph is shifted by an offset that is the number of qubits in the subsystem over 2. The three thin gray lines in (a) are $\log(3\times2^2),\log(3\times2^4),\log(3\times2^6)$. A straight line has been drawn in both (a) and (b) as a visual guide to show that the operator entanglement grows at most logarithmically.}
\label{OperatorEntanglement}
\end{figure}

\begin{figure}
\centering
\begin{subfigure}{0.48\linewidth}
    \centering
    \includegraphics[height=3cm, keepaspectratio]{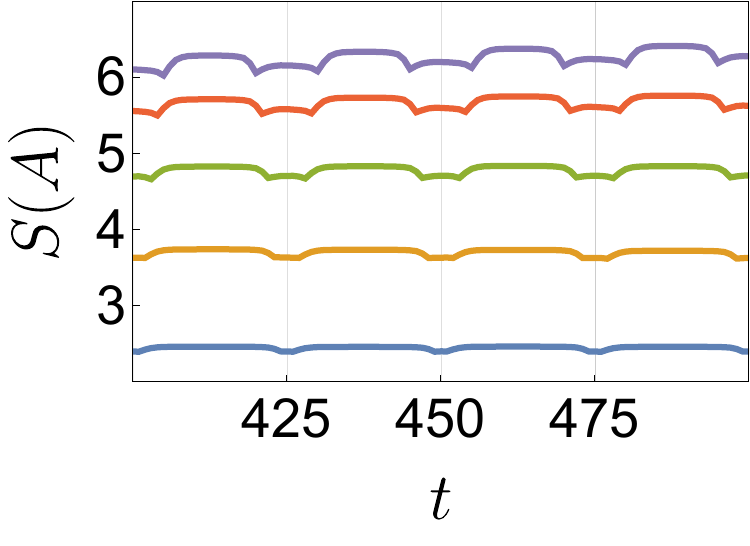}
    \caption{}
    \label{:sub_a}
\end{subfigure}
\hfill
\begin{subfigure}{0.48\linewidth}
    \centering
    \includegraphics[height=3cm, keepaspectratio,trim=1.3cm 0cm 0cm 0cm, clip]{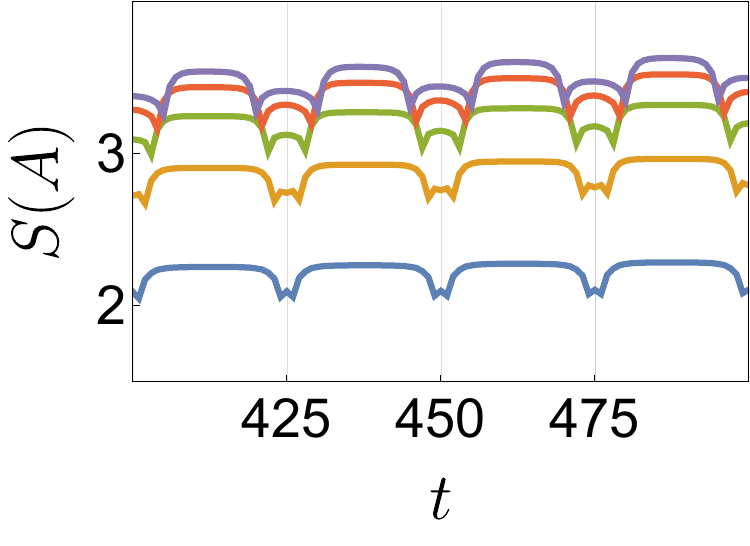}
    \caption{}
    \label{:sub_b}
\end{subfigure}
\includegraphics[scale=0.3]{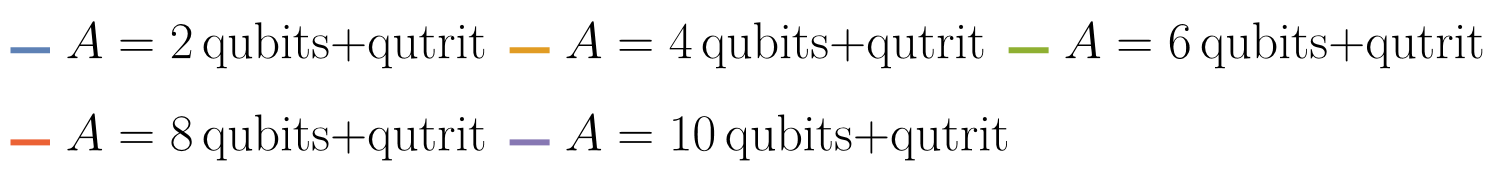}
\caption{These two plots (a) and (b) zoom into the little dips seen in the panels (a) and (b) in Fig.~\ref{OperatorEntanglement}, respectively.}
\label{OperatorEntanglementDips}
\end{figure}
In this subsection, we compute the operator entanglement of $\sigma^x$, initially placed on an odd-numbered site, undergoing Heisenberg time evolution. The operator entanglement for a generic operator $O$ in the restricted subspace \eqref{PauliStringsNonZeroCorrelatorSpecialForm} is simply the von Neumann entanglement entropy of the operator wavefunction \eqref{PauliStringZBasis}. It was first introduced in \cite{PhysRevA.63.040304,PhysRevA.76.032316,PhysRevB.79.184416} as a way to quantify the complexity and simulatability of operators by tensor networks and was subsequently used as a way to quantify information scrambling and quantum chaos \cite{Hosur2016,Nie_2019,Kudler-Flam2020,PhysRevResearch.3.033182,PhysRevB.104.214202,Goto2022,Goto2023,PhysRevResearch.6.023001}

Plots of the operator entanglement of $\sigma^x$ are shown in figure \ref{OperatorEntanglement} and figure \ref{OperatorEntanglementDips}. The plots (a) in figure \ref{OperatorEntanglement} and figure \ref{OperatorEntanglementDips} correspond to the semi-ergodic point while the plots (b) correspond to the non-semi-ergodic point. Figure \ref{OperatorEntanglement} show how the operator entanglement increases from $0$, as we started with a pure operator state $\sigma^x$, up to $\left(\frac{L}{2}\right)^2$ layers of unitaries applied. The horizontal axes is logarithmic and it is clear that in both cases, we see a operator entanglement growth that is at most logarithmic in time, with the operator entanglement for the semi-ergodic point saturating much earlier. We also see a larger operator entanglement at the semi-ergodic regime than at the non-semi-ergodic regime. This logarithmic growth is noteworthy because with the inclusion of the single-site gates, $U$ as defined in \eqref{DUMatrix} no longer produces an integrable circuit.

For each plot, we show the operator entanglement for different subsystem sizes, where each subsystem contains the single qutrit as well as an equal number of qubits on both sides of that qutrit at that instant in time. Since the qubits are moving relative to the qutrit, the qubits that are included in the subsystem changes with time even though the size of the subsystem remains fixed in time. Note that the horizontal axes for the top two plots in figure \ref{OperatorEntanglement} is shifted for each graph by the number of qubits in the subsystem divided by two. By performing this shift in the horizontal axes, the graphs of operator entanglement for different subsystem sizes lie on top of each other at early times. This can easily be understood from the scattering diagram fig. \ref{QutritQubitScatteringDiagram}. At time $t=\frac{l_A}{2}$, where $t$ layers of unitaries have been applied, the qutrit is entangled only with the $\frac{l_A}{2}$ qubits on its right, with which it has just scattered. Therefore, up until this $t=\frac{l_A}{2}$, all the entanglement in the system is contained within subsystem $A$ and the operator entanglement remains zero. At time $t = \frac{l_A}{2}+1$, the very first qubit that scattered with the qutrit is located immediately outside of the subsystem as the qutrit is now entangled with $\frac{l_A}{2}+1$ qubits on its right, and so at $t=\frac{l_A}{2}+1$, the operator entanglement becomes non-zero for the first time.

In the semi-ergodic case, the operator entanglement initially increases to a value of about $\ln 3$ which is the maximal amount of entanglement a qutrit can have. When $t\leq \frac{L}{2}$, the only entanglement present in the system is between the qutrit and the $t$ qubits it has scattered with. The operator entanglement, up to the horizontal offset, remains the same for all subsystem sizes, until $t+\frac{l_A}{2}-1=\frac{L}{2}$, when the very first qubit that scattered with the qutrit has re-entered the subsystem. This occurs at different times for different subsystem sizes so the operator entanglement will be subsystem size dependent at this time.
This causes the operator entanglement to increase above $\ln 3$, and occurs at different times for different subsystem sizes. The operator entanglement in the non-semi-ergodic case behaves similarly, but without the initial saturation at $\ln 3$.

Figure \ref{OperatorEntanglementDips} zoom in on the tiny bumps seen in figure \ref{OperatorEntanglement}. They occur at time steps of $L/2$ which is when the qutrit has scattered with all $L/2$ qubits an equal number of times. 

The most remarkable feature is a slow long-time growth of operator entropy. Even in the semi-ergodic regime we note a very slow growth of operator entanglement entropy 
until the ultimate plateau is being approached which is the logarithm of the
sub-system dimension. Based on extensive numerical data we conjecture the entanglement entropy to grow no faster than logarithmically in time.
This is very similar to the behavior of disordered qubit chains in the many-body-localized regime~\cite{znidaric2008,Pollmann2012,abanin13}, however, the mechanism here still needs to be explored.

\subsection{Operator Size}
We define the operator size for a Pauli string to be the number of non-identities appearing in that string. If we expand an operator in a basis of Pauli strings $\mathcal{P}$,
\begin{equation}
    O = \sum_{\mathcal{P}} c_\mathcal{P} \mathcal{P}
\end{equation}
define the sum of amplitudes squared for a given size $l$ as
\begin{equation}\label{SumOfAmplitudes}
    S_l = \sum_{|\mathcal{P}|=l} |c_\mathcal{P}|^2
\end{equation}

\begin{figure}
    \centering    
    \begin{subfigure}{0.48\linewidth}
    \centering
\includegraphics[height=3cm, keepaspectratio]{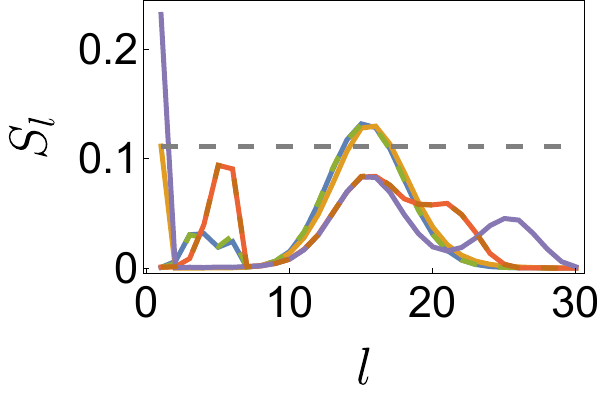}
    \caption{}
    \label{:sub_a}
\end{subfigure}
\hfill
\begin{subfigure}{0.48\linewidth}
    \centering
\includegraphics[height=3cm, keepaspectratio,trim=0.9cm 0cm 0cm 0cm, clip]{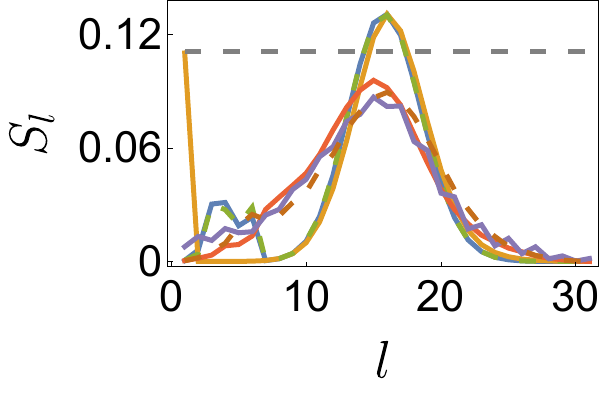}
    \caption{}
    \label{:sub_b}
\end{subfigure}
     \includegraphics[scale=0.48]{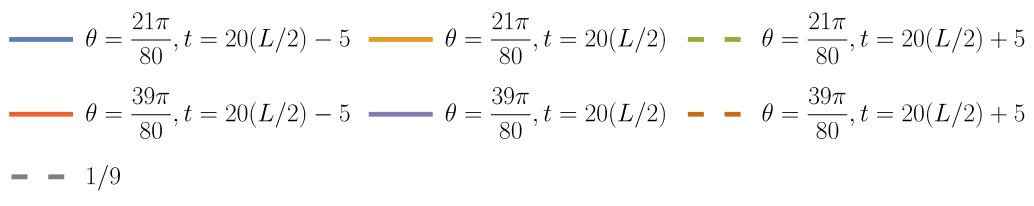}
    \caption{Plots of operator size distributions $S_l$ \eqref{SumOfAmplitudes} for a given size $l$ against the operator size $l$. The left (a) and right (b) plots corresponds to $L=58$ and $L=60$ respectively. For each plot, we show the plots for both $\theta=\frac{21\pi}{80}$ and $\theta=\frac{39\pi}{80}$, the semi-ergodic and non-semi-ergodic points respectively. We show the distribution of $S_l$ at a time step that is a multiple of $L/2$, as well as time steps slightly before and after this multiple of $L/2$.}
    \label{OperatorSize_ComparingDifferentTheta}
\end{figure}
Plots of such operator size distributions $S_l$ defined in \eqref{SumOfAmplitudes} are shown in figure \ref{OperatorSize_ComparingDifferentTheta}. We show both $L=58$ and $L=60$ because of the dependence on the parity of $L/2$ of the correlation functions which are directly related to the operator size. For both semi-ergodic and non-semi-ergodic values of $\theta$ when $L/2$ is odd, and for the semi-ergodic value of $\theta=21\pi/80$ when $L/2$ is even, we see a sharp peak in the operator size distribution at $l=1$, in addition to a broad peak of the operator size at larger values. This is to be expected because as we observed earlier, the auto-correlation function has a non-trivial (non-small) value at time steps that are multiples of $L/2$. For these choices of $L$ and $\theta$, we note that the operator size distribution $\Delta t$ time steps before and after a multiple of $L/2$ almost lie on top of each other.

\section{Conclusion and Discussion}
We considered a brickwork dual-unitary circuit at finite width, away from the analytically tractable regime. The two-site dual-unitary gates were chosen such that the two-point correlation functions between single-site traceless operators exhibit ergodic behavior along one light ray and non-ergodic behavior along the other light ray. It turned out that under Heisenberg evolution by this circuit, a single-site Pauli matrix located along the ergodic direction of the dual-unitary circuit evolves within a (exponentially) restricted subspace of operators. Time evolution in this restricted subspace of operators can be thought of as a single qutrit, which encodes a single Pauli operator state, scattering with $L/2$ qubits sequentially, where $L$ is the total system size. The time evolution of the corresponding operator wavefunction is given by an iterative multiplication by a six-by-six matrix which describes the scattering process between the qutrit and a single qubit. At every time step, only a single qubit is involved in the scattering while the rest of the qubits move along unaffected. This allows our exact numerics to reach relatively large system sizes (up to $L\approx 60$). Furthermore, we exploit a hidden symmetry of the scattering matrix to rewrite the correlation functions in terms of powers of products of SO(3) matrices. Assuming this sum and product of matrices approaches a Haar average, the auto-correlation function is predicted to be $1/3$ at late times that are multiples of $L/2$, where the qutrit has scattered with every single qubit an equal number of times.

The most interesting results are the dynamics of the operator entanglement and operator size distribution under this semi-ergodic dynamics. Despite being non-integrable, we observe at most logarithmic growth of the operator entanglement in time, contrary to prior expectations. The operator size distribution also becomes bimodal around time steps that are multiples of $L/2$, exhibiting the usual peak around large, complicated operators, but also a peak around smaller, simpler operators. This is a unique feature of semi-ergodic dynamics that sits between chaotic and free or integrable theories.

The most obvious extension of this work is to consider semi-ergodic dynamics in other generalizations of dual-unitary circuits, such as ternary circuits in higher dimensions \cite{PhysRevLett.130.090601}, or triunitary circuits in one spatial dimension \cite{PhysRevResearch.3.043046}. While our semi-ergodic dual-unitary circuit was not sufficiently complex for demonstrating quantum advantage, some of these extensions might be. For example, single-site correlations in triunitary circuits propagate along three directions instead of two. Perhaps by having a single non-ergodic direction and two ergodic ones, the dynamics would be tilted more towards chaotic behavior, potentially making the dynamics difficult to simulate with classical techniques at late times while still having non-trivial signals at late times due to the presence of a single non-ergodic channel. 

It would also be interesting to realize semi-ergodic dynamics in Hamiltonians systems that have bimodal operator size distributions, and to have a more systematic understanding as to when semi-ergodic dynamics might arise.

\begin{acknowledgments}
We acknowledge the support by European Research Council (ERC) through Advanced grant QUEST (Grant Agreement No. 101096208), and Slovenian Research and Innovation agency (ARIS) through the Program P1-0402 and Grant N1-0368.
We thank Sergey Filippov
and Joseph Tindall for fruitful discussions, and in particular, Joseph Tindall
 for providing the codes and advice for benchmarking the results against tensor network simulations.
 
\end{acknowledgments}

\nocite{*}
\appendix
\section{Correlation Functions}
In the main text, we focused on the auto-correlation functions. In this appendix, we show results pertaining to the off-diagonal correlation functions $C_{YX}$ and $C_{ZX}$. 
\begin{figure}
\centering
\begin{subfigure}{0.42\columnwidth}
    \centering
    \includegraphics[height=3.8cm, keepaspectratio]{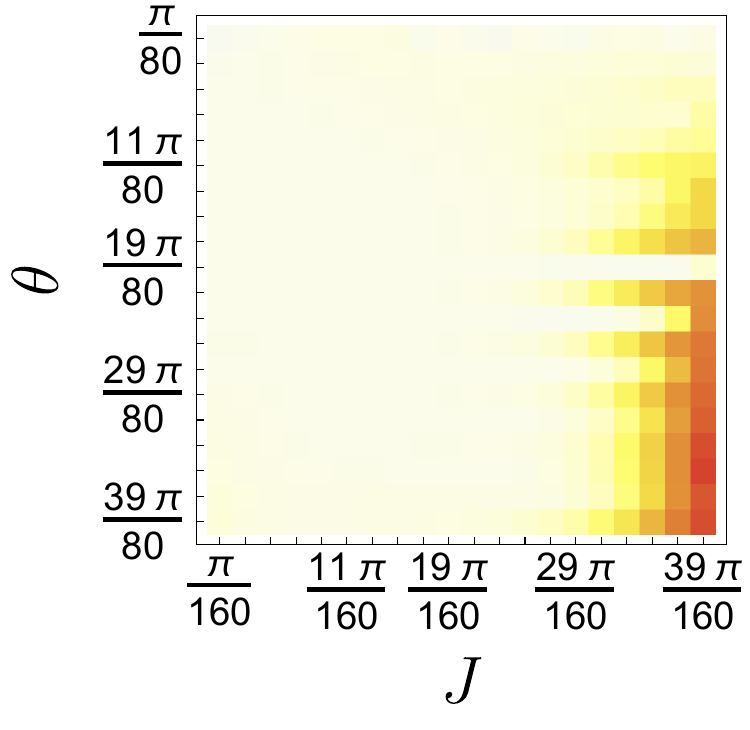}
    \caption{}
    \label{offDiagonalCorrelator:sub_a}
\end{subfigure}
\hfill
\begin{subfigure}{0.42\columnwidth}
    \centering
    \includegraphics[height=3.8cm, keepaspectratio,trim=1cm 0cm 0cm 0cm, clip]{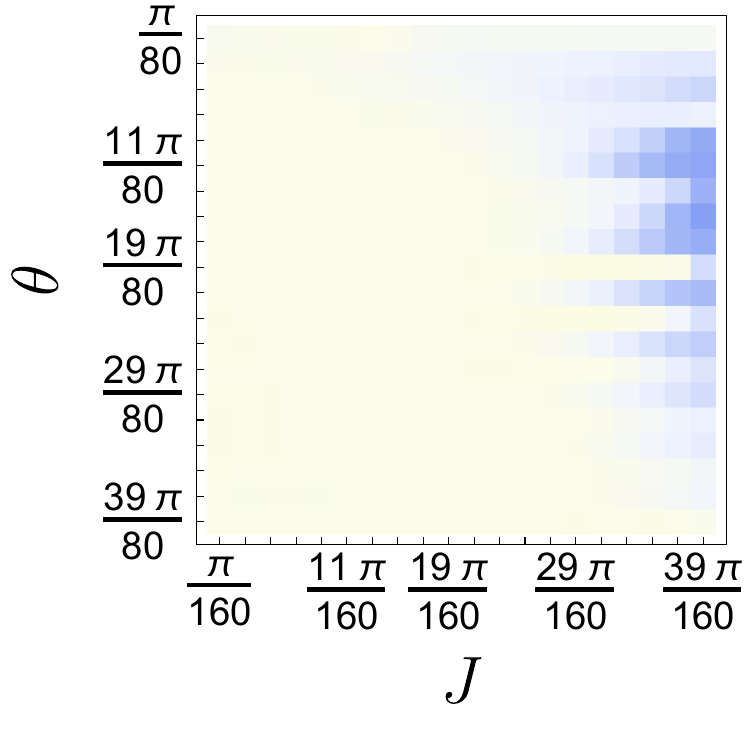}
    \caption{}
    \label{offDiagonalCorrelator:sub_b}
\end{subfigure}
\raisebox{1.5cm}{
\includegraphics[scale=0.5]{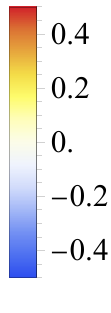}}
\caption{A heatmap of the time-averaged auto-correlator \eqref{TimeAveragedCorrelator}, (a)  $ \overline{C_{YX}}((\frac{L}{2})^2)$ and (b) $\overline{C_{ZX}}((\frac{L}{2})^2)$, evaluated at a time extensive in the system size $L$ which we take here to be $L=48$. The vertical axis is $\theta$ which parametrizes the single-site unitary $v_-$ \eqref{SingleSiteErgodic} while the horizontal axis is $J$ which parametrizes the entangling power of the two-site unitary.}
\label{offDiagonalCorrelator_HeatMap}
\end{figure}

Not much difference is observed in the heat maps for different values of system size $L$ so we only show the heat map for a single value of $L$ in figure \ref{offDiagonalCorrelator_HeatMap}. For most points in the heat map, the time-averaged correlation functions have come close to the Haar-averaged value of $0$ \eqref{HaarAveragePrediction}. Nevertheless, we still consider the previous point $J=\frac{11\pi}{160}$, $\theta=\frac{39\pi}{80}$ as non-semi-ergodic because for that choice of parameters, the auto-correlator has not approached the Haar-averaged value yet.

\begin{figure}
\centering
\begin{subfigure}{\columnwidth}
    \centering
    \includegraphics[width=\columnwidth]{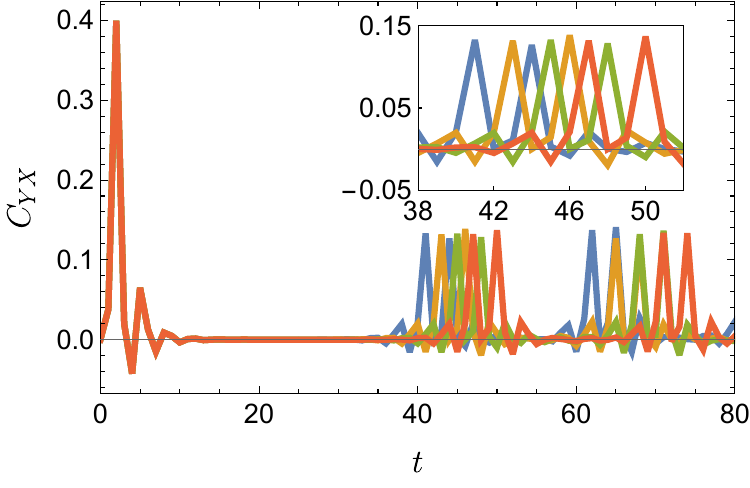}
    \caption{}
    \label{YXcorrelator:sub_a}
\end{subfigure}
\hfill
\begin{subfigure}{\columnwidth}
    \centering
    \includegraphics[width=\columnwidth]{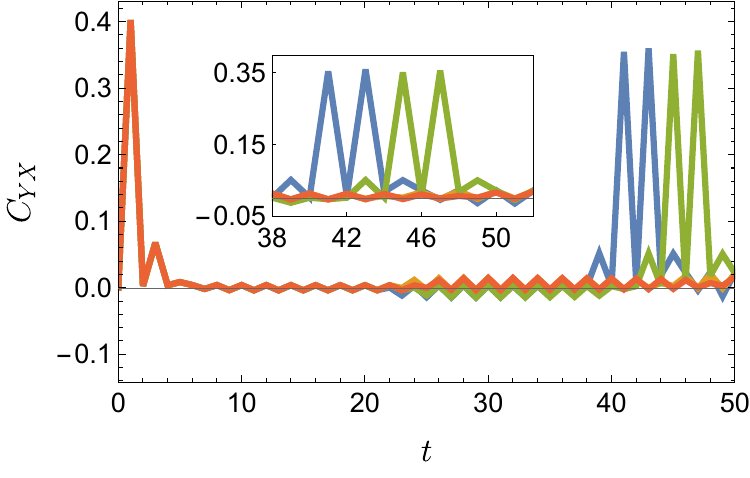}
    \caption{}
    \label{YXcorrelator:sub_b}
\end{subfigure}
\includegraphics[width=0.8\linewidth]{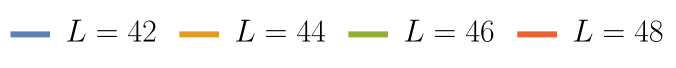}
\caption{(Main Plots) Plots of the correlation functions $C_{YX}$ against the number of single layers of unitaries for $J=\frac{11\pi}{160}$,$\psi = \frac{1+\sqrt{5}}{2}\frac{\pi}{2}$, $\phi = (1+\sqrt{2})\frac{\pi}{2}$ and various system sizes $L$. The top plot, (a), corresponds to a semi-ergodic point $\theta=\frac{21\pi}{80}$, while the bottom plot, (b), corresponds to a non-semi-ergodic point $\theta=\frac{39\pi}{80}$. (Insets) The insets show are the same plots but zoomed in around the spikes.}
\label{YXcorrelator}
\end{figure}

\begin{figure}
\centering
\begin{subfigure}{\columnwidth}
    \centering
    \includegraphics[width=\columnwidth]{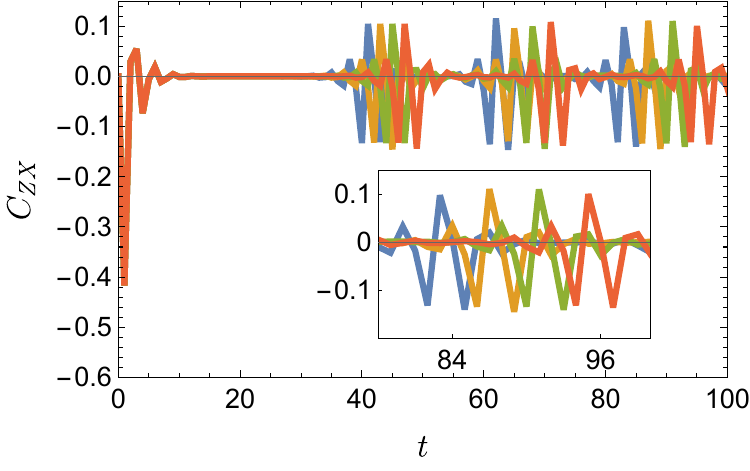}
    \caption{}
    \label{:sub_a}
\end{subfigure}
\hfill
\begin{subfigure}{\columnwidth}
    \centering
    \includegraphics[width=\columnwidth]{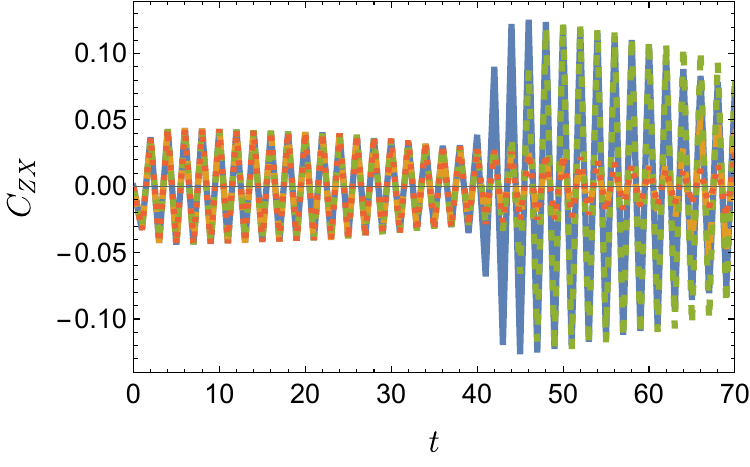}
    \caption{}
    \label{:sub_b}
\end{subfigure}
\includegraphics[width=0.8\linewidth]{CorrelatorPlots/YXcorrelator_Legend.png}
\caption{(Main Plots) Plots of the correlation functions $C_{ZX}$ against the number of single layers of unitaries for $J=\frac{11\pi}{160}$, $\psi = \frac{1+\sqrt{5}}{2}\frac{\pi}{2}$, $\phi = (1+\sqrt{2})\frac{\pi}{2}$ and various system sizes $L$. The top plot, (a), corresponds to a semi-ergodic point $\theta=\frac{21\pi}{80}$ while the bottom plot, (b), corresponds to a non-semi-ergodic point $\theta=\frac{39\pi}{80}$. (Insets) The insets show are the same plots but zoomed in around the spikes.}
\label{ZXcorrelator}
\end{figure}

Plots of the $C_{YX}$ and $C_{ZX}$ correlators are shown in figures \ref{YXcorrelator} and \ref{ZXcorrelator}. For both plots in figure \ref{YXcorrelator} and the first plot in figure \ref{ZXcorrelator}, we see spikes around, but not at, times that are multiples of $L/2$. This is to be expected because if we have a sizeable operator weight on $\sigma^x$ at times that are multiples of $L/2$, there must be a comparable operator weight for operators that are connected to $\sigma^x$ by a few layers of unitaries before and after a time that is a multiple of $L/2$. In particular, for the time step immediately before and after a multiple of $L/2$, some of the operators of support $1$ and $2$ should have non-trivial operator weight because they are connected to $\sigma^x$ by a single layer of unitaries.

\section{Operator Size}

\begin{figure}
    \centering    
    \begin{subfigure}{0.48\linewidth}
    \centering
\includegraphics[height=3cm, keepaspectratio]{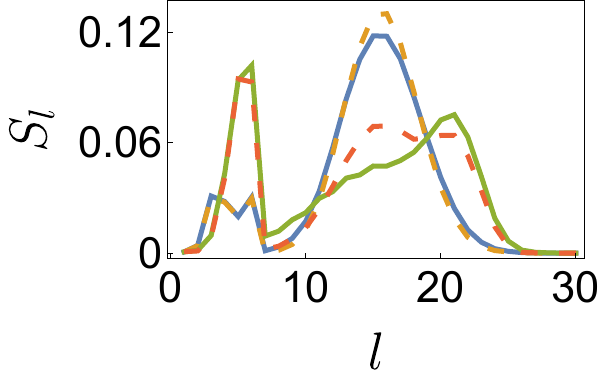}
    \caption{}
    \label{:sub_a}
\end{subfigure}
\hfill
\begin{subfigure}{0.48\linewidth}
    \centering
\includegraphics[height=3cm, keepaspectratio,trim=0.9cm 0cm 0cm 0cm, clip]{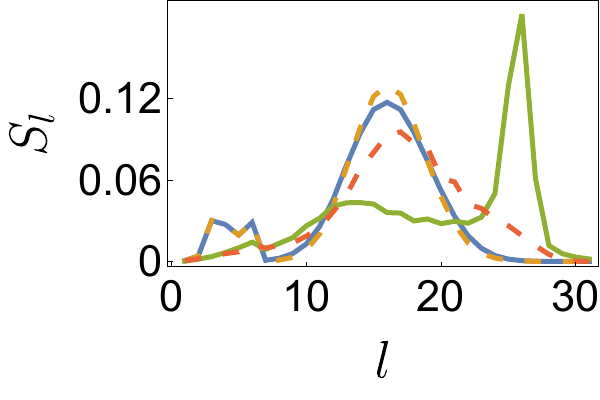}
    \caption{}
    \label{:sub_b}
\end{subfigure}
    \includegraphics[scale=0.5]{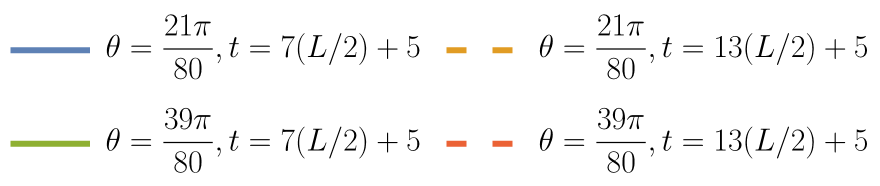}
    \caption{Plots of the sum of amplitudes $S_l$ \eqref{SumOfAmplitudes} for a given size $l$ against the operator size $l$. The left (a) and right (b) plots corresponds to $L=58$ and $L=60$ respetively. For each plot, we show the plots for both $\theta=\frac{21\pi}{80}$ and $\theta=\frac{39\pi}{80}$, the semi-ergodic and non-semi-ergodic points respectively. For each value of $\theta$, we show the distribution of $S_l$ at two different times that are shifted from different multiples of $L/2$ by the same offset $\Delta t=5$.}
    \label{OperatorSize_ComparingDifferentTimeSteps}
\end{figure}

In this section, we show additional plots for the operator size distribution. In figure \ref{OperatorSize_ComparingDifferentTimeSteps}, we show the operator size distribution for both odd $L/2$ and even $L/2$. Within each plot, we look at both the semi-ergodic point and non-semi-ergodic point, at times that are offsets of different multiples of $L/2$ with the same offset. For the semi-ergodic value of $\theta$ , we see that the bump in the operator size distribution at small operator sizes at different multiples of $L/2$ lie on top of each other. This is not always true for the non-semi-ergodic case. This suggests some kind of time translation symmetry for the peak in small operator sizes in the semi-ergodic regime.

\bibliography{main}

\end{document}